\documentclass[11pt]{article}
\usepackage[utf8]{inputenc}
\usepackage[T1]{fontenc}
\usepackage{authblk}
\usepackage[superscript]{cite}
\usepackage{multicol}
\usepackage{amsmath,amssymb,amsfonts,amsthm}
\usepackage[top=1in,bottom=1in,left=1in,right=1in]{geometry}
\usepackage{xspace,graphicx}
\theoremstyle{definition}

\theoremstyle{remark}

\catcode`,\active

\catcode`\,12

\newcommand*\pFqskip{8mu}
\catcode`,\active
\newcommand*\pFq{\begingroup
        \catcode`\,\active
        \def ,{\mskip\pFqskip\relax}%
        \dopFq
}
\catcode`\,12
\def\dopFq#1#2#3#4#5{%
        {}_{#1}F_{#2}\biggl[\genfrac..{0pt}{}{#3}{#4};#5\biggr]%
        \endgroup}

\renewcommand{\phi}{\varphi}

\newcommand{\under}[1]{_{#1}}%For references section%
\numberwithin{equation}{section}
\title{The algebra of dual $-1$ Hahn polynomials and the Clebsch-Gordan problem of $sl_{-1}(2)$}
\author[1]{Vincent X. Genest\thanks{genestvi@crm.umontreal.ca}}
\author[1]{Luc Vinet\thanks{luc.vinet@umontreal.ca}}
\author[2]{Alexei Zhedanov\thanks{zhedanov@kinetic.ac.donetsk.ua}}
\affil[1]{Centre de recherches math\'ematiques, Universit\'e de Montr\'eal, C.P. 6128, Succursale Centre-ville, Montr\'eal, Qu\'ebec, Canada H3C 3J7}
\affil[2]{Donetsk Institute for Physics and Technology, Donetsk 83114, Ukraine}
\begin{document}
\maketitle
\hrule
\begin{abstract}
\noindent
The algebra $\mathcal{H}$ of the dual $-1$ Hahn polynomials is derived and shown to arise in the Clebsch-Gordan problem of $sl_{-1}(2)$. The dual $-1$ Hahn polynomials are the bispectral polynomials of a discrete argument obtained from the $q\rightarrow -1$ limit of the dual $q$-Hahn polynomials. The Hopf algebra $sl_{-1}(2)$ has four generators including an involution, it is also a $q\rightarrow -1$ limit of the quantum algebra $sl_{q}(2)$ and furthermore, the dynamical algebra of the parabose oscillator. The algebra $\mathcal{H}$, a two-parameter generalization of $\mathfrak{u}(2)$ with an involution as additional generator, is first derived from the recurrence relation of the -1 Hahn polynomials. It is then shown that $\mathcal{H}$ can be realized in terms of the generators of two added $sl_{-1}(2)$ algebras, so that the Clebsch-Gordan coefficients of $sl_{-1}(2)$ are dual -1 Hahn polynomials. An irreducible representation of $\mathcal{H}$ involving five-diagonal matrices and connected to the difference equation of the dual $-1$ Hahn polynomials is constructed.
\end{abstract}
\hrule
\section*{Introduction}
\normalsize
\thispagestyle{empty}
The algebra $sl_{-1}(2)$ has been proposed\cite{Vinet-2011} as a $q\rightarrow -1$ limit of the  $sl_{q}(2)$ algebra. It is a Hopf algebra with four generators, including an involution, defined by relations involving both commutators and anti-commutators. This algebra is also the dynamical algebra of a parabosonic oscillator\cite{Green-1953,Mukunda-1980,Mukunda-1981,Rosenblum-1994} .

Recently, a breakthrough in the theory of orthogonal polynomials has been realized with the discovery of a series of classical orthogonal polynomials which are eigenfunctions of continuous or discrete  Dunkl operators defined using reflections\cite{Genest-2012,VZhedanov-2011,Vinet-2012,Vinet-2011-3,Vinet-2011-2,Zhedanov-2011} . These polynomials are referred to as $-1$ polynomials since they arise as $q\rightarrow -1$ limits of $q$-orthogonal polynomials of the Askey scheme. At the top of the discrete variable branch of these $q=-1$ polynomials lie the Bannai-Ito polynomials and their kernel partners, the complementary Bannai-Ito polynomials. Both sets depend on four parameters and are expressible in terms of Wilson polynomials\cite{Ito-1984,Koekoek-2010,Vinet-2012} . The Bannai-Ito polynomials possess the Leonard duality property\cite{Terwilliger-2001} , which in fact led to their original discovery\cite{Ito-1984} . Moreover, an algebraic interpretation of these polynomials has been given in terms of the Bannai-Ito algebra, which is a Jordan algebra\cite{Tsujimoto-2011} . In contradistinction to the situation with the Bannai-Ito polynomials, the complementary Bannai-Ito polynomials and their descendants, the dual $-1$ Hahn polynomials, are bi-spectral (i.e. they obey both a recurrence relation and a difference equation) but they fall outside the scope of the Leonard duality. Moreover, their algebraic interpretation is lacking.

In the present work, we derive the algebra $\mathcal{H}$ of the dual $-1$ Hahn polynomials and show that it arises as the hidden symmetry algebra of the Clebsch-Gordan problem of $sl_{-1}(2)$. It is already known\cite{Vinet-2011} that the dual $-1$ Hahn polynomials occur as Clebsch-Gordan coefficients of $sl_{-1}(2)$. Here we recover this result by showing how $\mathcal{H}$ is realized by generators of the coproduct of two $sl_{-1}(2)$ algebras. The algebra $\mathcal{H}$ turns out to be an extension of $\mathfrak{u}(2)$ through the addition of an involution as a generator. We study its finite-dimensional irreducible representations in two bases each diagonalizing a different operator.

The paper is divided as follows. In section 1, we recall basic results on the $sl_{-1}(2)$ algebra and the dual $-1$ Hahn polynomials. In section 2, we obtain the algebra $\mathcal{H}$ of the dual $-1$ Hahn polynomials in a specific representation by using the recurrence relation operator and the spectrum of the difference equation. In section 3, we investigate the Clebsch-Gordan problem for $sl_{-1}(2)$ and show that $\mathcal{H}$ appears as the associated algebra. In section 4, an irreducible representation of $\mathcal{H}$ which is ''dual'' to the one constructed in section 2 is shown to involve five-diagonal matrices. We conclude by discussing another presentation of $\mathcal{H}$ and its relation to the algebras proposed\cite{VDJ-2011,VDJ-2012} in the context of finite oscillator models.
\section{$sl_{-1}(2)$ and dual $-1$ Hahn polynomials}
\subsection{The algebra $sl_{-1}(2)$}
The  Hopf algebra $sl_{-1}(2)$ is generated by four operators $A_0$, $A_{+}$, $A_{-}$ and $R$ obeying the relations\cite{Vinet-2011} 
\begin{align}
\label{DefiningRelations}
[A_0,A_{\pm}]=\pm A_{\pm},\quad [A_0,R]=0,\quad\{A_{+},A_{-}\}=2A_{0},\quad \{A_{\pm},R\}=0,
\end{align}
where $[a,b]=ab-ba$ and $\{a,b\}=ab+ba$. The operator $R$ is an involution, that is
$
R^2=\mathbb{Id},
$
where $\mathbb{Id}$ represents the identity operator. The algebra has the  following Casimir operator:
\begin{align}
\label{Casimir}
Q=A_{+}A_{-}R-A_{0}R+(1/2)R,
\end{align}
which commutes with all generators. In view of the defining relations \eqref{DefiningRelations}, it is clear that $sl_{-1}(2)$ has a ladder representation. Let $\mu$ be a non-negative real number and let $\epsilon$ be a parameter taking the values $\epsilon=\pm 1$. Consider the infinite-dimensional vector space $(\epsilon,\mu)$ spanned by the basis vectors $e^{(\epsilon,\mu)}_{n}$, $n\in \mathbb{N}$, and endowed with the actions
\begin{align}
&A_0\,e_{n}^{(\epsilon,\mu)}=(n+\mu+1/2)\,e_{n}^{(\epsilon,\mu)}, && R\,e_{n}^{(\epsilon,\mu)}=\epsilon (-1)^{n}\,e_{n}^{(\epsilon,\mu)},\nonumber\\
\label{Product-1}
&A_{+}\,e_{n}^{(\epsilon,\mu)}=\surd [n+1]_{\mu}\,e_{n+1}^{(\epsilon,\mu)}, &&A_{-}\,e_{n}^{(\epsilon,\mu)}=\surd [n]_{\mu}\,e_{n-1}^{(\epsilon,\mu)},
\end{align}
where $[n]_{\mu}$ denotes the $\mu$-number
\begin{align}
\label{MuNombre}
[n]_{\mu}=n+\mu(1-(-1)^{n}).
\end{align}
It is readily checked that \eqref{Product-1} defines an irreducible $sl_{-1}(2)$-module. As expected from Schur's lemma, the Casimir operator $Q$ acts on $(\epsilon,\mu)$ as a multiple of the identity:
\begin{align}
\label{CasimirAction}
Q\,e_{n}^{(\epsilon,\mu)}=-\epsilon\mu\,e_{n}^{(\epsilon,\mu)}.
\end{align}
On the space $(\epsilon,\mu)$, $sl_{-1}(2)$ is equivalent to the dynamical algebra of a parabosonic oscillator. This assertion stems from the following  observation. One has
\begin{align}
[A_{-},A_{+}]=\{A_{-},A_{+}\}-2A_{+}A_{-}=2A_0-2A_{+}A_{-}.
\end{align}
With the use of \eqref{Casimir} and \eqref{CasimirAction}, one finds
\begin{align}
\label{ActionIrreps}
[A_{-},A_{+}]=1+2\epsilon\mu R,
\end{align}
where the relation is understood to be on the space $(\epsilon,\mu)$. The operators $A_{\pm}$ satisfying the commutation relation \eqref{ActionIrreps}, together with the involution $R$ obeying the relations $R^2=\mathbb{Id}$ and $\{R,A_{\pm}\}=0$, define the parabosonic oscillator algebra\cite{Mukunda-1980,Rosenblum-1994,Mukunda-1981,Vinet-2011} .

The $sl_{-1}(2)$ algebra possesses a non-trivial addition rule, or coproduct\cite{Daska-2000,Vinet-2011} . Let $(\epsilon_a,\mu_a)$ and $(\epsilon_b,\mu_b)$ be two $sl_{-1}(2)$-modules. A third module is obtained by endowing the tensor product space $(\epsilon_a,\mu_a)\otimes(\epsilon_b,\mu_b)$ with the actions
\begin{align}
A_0(u\otimes v)&=(A_0u)\otimes v+u\otimes(A_0v),\nonumber\\
A_{\pm}(u\otimes v)&=(A_{\pm}u)\otimes (Rv)+u\otimes (A_{\pm}v),\\
R(u\otimes v)&=(Ru)\otimes(Rv),\nonumber
\end{align}
where $u\in(\epsilon_a,\mu_a)$ and $v\in (\epsilon_b,\mu_b)$. This addition rule for $sl_{-1}(2)$ can also be presented in the following way. Let $\{A_{0},A_{\pm},R_{a}\}$ and $\{B_{0},B_{\pm},R_{b}\}$ be two mutually commuting sets of $sl_{-1}(2)$ generators and denote the corresponding algebras by $\mathcal{A}$ and $\mathcal{B}$. A third algebra, denoted $\mathcal{C}=\mathcal{A}\oplus\mathcal{B}$, is obtained by defining
\begin{align}
C_0=A_0+B_0, && C_{\pm}=A_{\pm}R_{b}+B_{\pm},&&& R_{c}=R_{a}R_{b}.
\end{align}
It is elementary to verify that the generators $C_0$, $C_{\pm}$ and $R_{c}$ obey the defining relations \eqref{DefiningRelations} of $sl_{-1}(2)$. The Casimir operator of the resulting algebra
\begin{align}
Q_{ab}=C_{+}C_{-}R_{c}-C_{0}R_{c}+(1/2)R_{c},
\end{align}
may be cast in the form
\begin{align}
Q_{ab}=\left(A_{-}B_{+}-A_{+}B_{-}\right)R_{a}-(1/2)R_{a}R_{b}+Q_{a}R_{b}+Q_{b}R_{a},
\end{align}
where $Q_{i}$, $i\in \{a,b\}$, are the Casimir operators of the algebras $\mathcal{A}$ and $\mathcal{B}$.
\subsection{Dual $-1$ Hahn polynomials}
The dual $-1$ Hahn polynomials $Q_{n}(x;\alpha,\beta,N)$ have been introduced and investigated\cite{VZhedanov-2011} as limits of the dual $q$-Hahn polynomials\cite{Koekoek-2010} when $q\rightarrow -1$. We here recall their basic properties. The monic dual $-1$ Hahn polynomials obey the recurrence relation
\begin{align}
Q_{n+1}(x;\alpha,\beta,N)+b_{n}Q_{n}(x;\alpha,\beta,N)+u_{n}Q_{n-1}(x;\alpha,\beta,N)=xQ_{n}(x;\alpha,\beta,N).
\end{align}
The recurrence coefficients are expressed in terms of $\mu$-numbers \eqref{MuNombre} as follows:
\begin{align}
\label{RecurrenceCoefficients}
b_{n}=
\begin{cases}
(-1)^{n+1}(2\xi+2\zeta)-1, & \text{$N$ even,}\\
(-1)^{n+1}(2\xi-2\zeta)-1, & \text{$N$ odd,}
\end{cases}\qquad
u_{n}=4[n]_{\xi}[N-n+1]_{\zeta},
\end{align}
where the parameters $\xi$ and $\zeta$ are given by
\begin{align}
\xi=
\begin{cases}
\frac{\beta-N-1}{2}, & \text{$N$ even},\\
\frac{\alpha}{2}, & \text{$N$ odd},
\end{cases}
\qquad
\zeta=
\begin{cases}
\frac{\alpha-N-1}{2}, & \text{$N$ even},\\
\frac{\beta}{2}, & \text{$N$ odd}.
\end{cases}
\end{align}
It is seen that the truncation conditions $u_0=0$ and $u_{N+1}=0$, necessary for finite orthogonal polynomials, are met. The positivity condition $u_{n}>0$ is equivalent to the condition $\alpha>N$ and $\beta>N$ in the case of even $N$. When $N$ is odd, two conditions are possible to ensure positivity; the relevant situation here is $\alpha>-1$ and $\beta>-1$. The dual $-1$ Hahn polynomials enjoy the orthogonality relation
\begin{align}
\sum_{s=0}^{N}\omega_{s}Q_{n}(x_s;\alpha,\beta,N)Q_{m}(x_s;\alpha,\beta,N)=v_{n} \delta_{nm}.
\end{align}
The grid and weight function are given by
\begin{align}
\label{grid}
x_{s}&=
\begin{cases}
(-1)^{s}(2s+1-\alpha-\beta), & \text{$N$ even},\\
(-1)^{s}(2s+1+\alpha+\beta), & \text{$N$ odd},
\end{cases}
\\
\omega_{2j+q}&=
\begin{cases}
(-1)^{j}\frac{(-N/2)_{j+q}}{j!}\frac{(1-\alpha/2)_{j}(1-\alpha/2-\beta/2)_{j}}{(1-\beta/2)_j(N/2+1-\alpha/2-\beta/2)_{j+q}}, & \text{$N$ even,}\\
(-1)^{j}\frac{(-(N-1)/2)_{j}}{j!}\frac{(1/2+\alpha/2)_{j+q}(1+\alpha/2+\beta/2)_j}{(1/2+\beta/2)_{j+q}(N/2+3/2+\alpha/2+\beta/2)_j}, & \text{$N$ odd,}
\end{cases}
\end{align}
where $q\in\{0,1\}$ and where $(a)_n=(a)(a+1)\cdots(a+n-1)$ stands for the Pochhammer symbol. The normalization coefficients $v_n$ take the form
\small
\begin{equation}
v_{n}=
\begin{cases}
(-1)^{q}(16)^{2j+q}j!(1-\alpha/2)_{j}(-N/2)_{j+q}(\beta/2-N/2)_{j+q}\left(\frac{(1-(\alpha+\beta)/2)_{N/2}}{(1-\beta/2)_{N/2}}\right),& \text{$N$ even,}\\
(-1)^{q}(16)^{2j+q}j!(1/2+\alpha/2)_{j+q}(1/2-N/2)_{j}(-\beta/2-N/2)_{j+q}\left(\frac{(1+(\alpha+\beta)/2)_{\lceil N/2\rceil}}{((\beta+1)/2)_{\lceil N/2\rceil}}\right), & \text{$N$ odd},
\end{cases}
\end{equation}
\normalsize
where $\lceil x\rceil=\lfloor x\rfloor+1 $ and $\lfloor x\rfloor$ denotes the integer part of $x$. The dual $-1$ Hahn polynomials admit a hypergeometric representation. Recall that the generalized hypergeometric function $_{p}F_{q}(z)$ is defined by the infinite series
\begin{align}
_{p}F_{q}
\left[
\begin{matrix}
a_1 & \cdots & a_{p} \\
b_1 & \cdots & b_{q}
\end{matrix};\;z\right]=\sum_{k}\frac{(a_1)_{k}\cdots (a_p)_{k}}{(b_1)_{k}\cdots (b_q)_{k}}\frac{z^{k}}{k!}.
\end{align}
When $N$ is even, one has\cite{VZhedanov-2011}
\begin{align}
Q_{2n}(x)&=\gamma_{n}^{(0)}\,\pFq{3}{2}{-n,\delta+\frac{x+1}{4},\delta-\frac{x+1}{4}}{-\frac{N}{2},1-\frac{\alpha}{2}}{1},\qquad\delta=1/2-\frac{\alpha+\beta}{4},\\
Q_{2n+1}(x)&=\gamma_{n}^{(1)}(x+1-\tau)\pFq{3}{2}{-n,\delta+\frac{x+1}{4},\delta-\frac{x+1}{4}}{1-\frac{N}{2},1-\frac{\alpha}{2}}{1},\quad\tau=2N+2-\alpha-\beta,
\end{align}
where $\gamma_{n}^{(0)}=16^{n}(-N/2)_{n}(1-\alpha/2)_{n}$ and $\gamma_{n}^{(1)}=16^{n}(1-N/2)_{n}(1-\alpha/2)_{n}$. When $N$ is odd, one rather has
\begin{align}
Q_{2n}(x)&=\phi_{n}^{(0)}
\pFq{3}{2}{-n,\eta+\frac{x+1}{4},\eta-\frac{x+1}{4}}{\frac{1-N}{2},\frac{\alpha+1}{2}}{1},\qquad
\eta=\frac{\alpha+\beta+2}{4},\\
Q_{2n+1}(x)&=\phi_{n}^{(1)}(x+1+\alpha-\beta)\,
\pFq{3}{2}{-n,\eta+\frac{x+1}{4},\eta-\frac{x+1}{4}}{\frac{1-N}{2},\frac{\alpha+3}{2}}{1},
\end{align}
where $\phi_{n}^{(0)}=16^{n}((1-N)/2)_{n}((\alpha+1)/2)_{n}$ and $\phi_{n}^{(1)}=16^{n}((1-N)/2)_{n}((\alpha+3)/2)_{n}$.

The dual $-1$ Hahn polynomials are bispectral but fall outside the scope of the Leonard duality. In point of fact, they satisfy a \emph{five-term} (instead of three-term) difference equation on the grid $x_{s}$. This equation is of the form\cite{VZhedanov-2011}:
\small
\begin{align}
A(s)Q_{n}(x_{s+2})+B(s)Q_{n}(x_{s+1})+C(s)Q_{n}(x_{s})+D(s)Q_{n}(x_{s-1})+E(s)Q_{n}(x_{s-2})=2n Q_{n}(x_{s}).
\end{align}
\normalsize
It is derived from the $q\rightarrow -1$ limit of the operator $L^2+2L$, where $L$ is the difference operator of the dual $q$-Hahn polynomials\cite{Koekoek-2010} . The expressions of the coefficients $A(s)$, $B(s)$, $C(s)$, $D(s)$ and $E(s)$ are known explicitly\cite{VZhedanov-2011} . In relation with this structure of the difference operator, we show  in Section 4 that the algebra $\mathcal{H}$ of the dual $-1$ Hahn polynomials admits an irreducible representation which involves five-diagonal matrices.
\section{The algebra $\mathcal{H}$ of $-1$ Hahn polynomials}
A specific realization of the algebra of the dual $-1$ Hahn polynomials is derived by examining the interplay between the recurrence and the difference operators. We consider the finite-dimensional vector space spanned by the basis elements $\psi_{n}$, $n\in \{0,\ldots,N\}$, and define the following operators:
\begin{align}
\label{Realization}
K_1\psi_{n}=n\psi_{n}, && 2K_2\psi_{n}=\psi_{n-1}+b_{n}\psi_{n}+u_{n}\psi_{n+1},
\end{align}
where $b_{n}$ and $u_{n}$ are as specified by \eqref{RecurrenceCoefficients}. Note that $\psi_{-1}$ and $\psi_{N+1}$ do not belong to the vector space so that the action of $K_2$ on the endpoint vectors $\psi_{0}$, $\psi_{N}$ is given by
\begin{align}
2K_2\psi_0=b_0\psi_{0}+u_{0}\psi_{1},\qquad 2K_2\psi_{N}=\psi_{N-1}+b_{N}\psi_{N}.
\end{align}
It is also necessary to introduce the parity operator $P$, which has the following realization in the basis $\psi_{n}$:
\begin{align}
P\psi_{n}=(-1)^{n}\psi_{n}.
\end{align}
It is seen that the operators $K_1$ and $K_2$, together with the parity operator $P$, are closed in frames of an algebra which we denote by $\mathcal{H}$. Indeed, a direct computation shows that the following relations hold:
\begin{align}
\label{First}
[K_1,P]=0, && \{K_2,P\}=-P-2\nu, &&& \{K_3,P\}=0,
\end{align}
\begin{align}
\label{Second}
[K_1,K_2]\equiv K_3, && [K_1,K_3]=K_2+\nu P+1/2,
\end{align}
\begin{align}
\label{Third}
[K_3,K_2]=4 K_1+4\nu K_1P-2\nu K_3P+\sigma P+\rho,
\end{align}
where the structure constants are
\begin{align}
\label{Para-Even}
\nu=(\alpha+\beta-2N-2)/2,\qquad \sigma=\alpha-\beta+2N(N+1-\beta),\qquad \rho=\beta-\alpha-2N,
\end{align}
when $N$ is even and
\begin{align}
\label{Para-Odd}
\nu=(\alpha-\beta)/2,\qquad \sigma=-(\alpha+\beta+2\alpha\beta+2N\alpha),\qquad \rho=\alpha-\beta-2N,
\end{align}
when $N$ is odd. The algebra $\mathcal{H}$, defined by the relations \eqref{First}, \eqref{Second} and \eqref{Third}, admits the Casimir operator
\begin{align}
\label{CasimirHAlgebra}
Q_{\mathcal{H}}=4K_1^2+K_2^2-K_3^2+K_2+2\rho K_1+2\nu P,
\end{align}
which commutes with all generators. The abstract algebra $\mathcal{H}$ is realized by the operators of the dual $-1$ Hahn polynomials as given by \eqref{Realization}. In this realization, the Casimir operator \eqref{CasimirHAlgebra}  takes the definite value
\begin{align}
\label{Casimir-Odd}
q_{\mathcal{H}}=\nu^2+2\nu-\sigma-\rho-\frac{1}{4}.
\end{align}
The picture can be summarized as follows. The algebra $\mathcal{H}$ has an irreducible representation of dimension $N+1$ for which the matrix representing $K_2$ is, up to a multiplicative constant, the Jacobi matrix of the monic dual $-1$ Hahn polynomials. By construction, one thus has
\small
\begin{align}
K_1=\mathrm{diag}(0,1,\cdots,N),
 &&
 K_2=\frac{1}{2}
\begin{pmatrix}
b_0 & 1 & & & \\
u_1 & b_1 & 1 & & \\
 & \ddots& \ddots &\ddots & \\
& & & & 1\\
& & & u_{N}& b_{N}
\end{pmatrix}, &&&
P=\mathrm{diag}(1,-1,1,\cdots,(-1)^{N}).
\end{align}
\normalsize
The representation is irreducible since $u_n$ is always non-zero. The transition matrix $S$ with matrix elements
\begin{align}
\label{Transition-Matrix}
S_{ij}=Q_{i}(x_{j};\alpha,\beta,N),
\end{align}
in the $\{\psi_{n}\}$ basis provides the similarity transformation diagonalizing $K_2$. Equivalently, the dual $-1$ Hahn polynomials are, up to factors, the overlap coefficients of the bases in which either $K_1$ or $K_2$ is diagonal. It is clear that in the basis in which $K_2$ is diagonal, the operators $P$ and $K_1$ will not be diagonal. In Section 4, the matrix elements of $P$ and $K_1$ in this basis will be  constructed from the commutation relations \eqref{First}, \eqref{Second} and \eqref{Third}. Unsurprisingly, the operator $K_1$ will be shown to be five-diagonal in this basis as expected from the form of the difference equation. We now turn to the Clebsch-Gordan problem.
\section{The Clebsch-Gordan problem}
The Clebsch-Gordan problem for $sl_{-1}(2)$ can be posited in the following way. We consider the $sl_{-1}(2)$-module $(\epsilon_a,\mu_a)\otimes(\epsilon_b,\mu_b)$ that we wish to decompose irreducibly. The basis vectors $e^{(\epsilon_a,\mu_a)}_{n_a}\otimes e^{(\epsilon_b,\mu_b)}_{n_b}$ of the direct product are characterized as eigenvectors of the operators
\begin{align}
\label{Label-1}
Q_{a},\;\; A_{0},\;\; R_{a}, \;\; Q_{b}, \;\; B_{0}, \;\; R_{b},
\end{align}
with eigenvalues
\begin{align}
-\epsilon_{a}\mu_{a},\;\; n_a+\mu_{a}+1/2, \;\; (-1)^{n_{a}}\epsilon_{a}, \;\; -\epsilon_{b}\mu_{b}, \;\; n_b+\mu_b+1/2, \;\; (-1)^{n_b}\epsilon_{b},
\end{align}
respectively. The irreducible modules in the decomposition will be spanned by the elements $e_{k}^{(\epsilon_{ab},\mu_{ab})}$ refered to as the coupled basis vectors. In each irreducible component, the (total) Casimir operator $Q_{ab}$ of the two added $sl_{-1}(2)$ algebras which reads
\begin{align}
\label{Casi}
Q_{ab}=\left(A_{-}B_{+}-A_{+}B_{-}\right)R_{a}-(1/2)R_{a}R_{b}+Q_{a}R_{b}+Q_{b}R_{a},
\end{align}
acts as a multiple of the identity:
\begin{align}
Q_{ab}=-\epsilon_{ab}\mu_{ab}\,\mathbb{Id}.
\end{align}
The coupled basis elements $e_{k}^{(\epsilon_{ab},\mu_{ab})}$are the eigenvectors of
\begin{align}
\label{Label-2}
Q_{ab},\;\;R_{a}R_{b},\;\; Q_{a},\;\; Q_{b},\;\; A_0+B_0,
\end{align}
with eigenvalues
\begin{align}
-\epsilon_{ab}\mu_{ab}, \;\; \epsilon_{ab},\;\; -\epsilon_{a}\mu_{a},\;\;-\epsilon_{b}\mu_{b},\;\;k+\mu_{a}+\mu_{b}+1,
\end{align}
respectively. The direct product basis is related to the coupled basis by a unitary transformation whose matrix elements are called Clebsch-Gordan coefficients. These overlap coefficients will be zero unless
\begin{align}
k&=n_{a}+n_{b}\equiv N.
\end{align}
We may hence write
\begin{align}
\label{Relation}
e_{N}^{(\epsilon_{ab},\mu_{ab})}=\sum_{n_a+n_b=N}C^{\mu_a\mu_b\mu_{ab}}_{n_an_bN}\; e_{n_a}^{(\epsilon_a\mu_a)}\otimes e_{n_b}^{(\epsilon_b\mu_b)},
\end{align}
where $C^{\mu_a\mu_b\mu_{ab}}_{n_an_bN}$ are the Clebsch-Gordan coefficients of $sl_{-1}(2)$.

The Clebsch-Gordan problem for $sl_{-1}(2)$ can be solved elegantly by examining the underlying symmetry algebra. It is first observed that for a given $N$, the following operators act as multiples of the identity operator on both sides of \eqref{Relation}:
\begin{align}
\label{Set}
\Lambda_1=2Q_{a},\qquad \Lambda_{2}=2Q_{b},\qquad \Lambda_{3}=R_{a}R_{b},\qquad \Lambda_{4}=A_0+B_0,
\end{align}
with multiples
\begin{align}
\lambda_{1}=-2\epsilon_{a}\mu_a,\qquad \lambda_{2}=-2\epsilon_{b}\mu_{b},\qquad \lambda_{3}=(-1)^{N}\epsilon_{a}\epsilon_{b},\qquad \lambda_{4}=\mu_a+\mu_b+N+1.
\end{align}
Let us now introduce the following operators
\begin{align}
\kappa_1=(A_0-B_0)/2,\qquad\kappa_2=\Lambda_{3}Q_{ab},\qquad r=R_{a}.
\end{align}
The Clebsch-Gordan problem for $sl_{-1}(2)$ is tantamount to finding the overlaps between the eigenvectors of $\kappa_1$ and $r$ and the eigenvectors of $\kappa_2$. The first set of eigenvectors correspond to the elements of the direct product basis (R.H.S. of \eqref{Relation}) since $\kappa_1$ and $r$ complement the set \eqref{Set} to give all the labelling operators \eqref{Label-1}. The second set is identified as should be to the coupled basis elements (L.H.S. of \eqref{Relation}) since only $Q_{ab}$ needs to be added to \eqref{Set} to recover the complete set of operators \eqref{Label-2} that are diagonal on the coupled vectors $e_{N}^{(\epsilon_{ab},\mu_{ab})}$; it will prove convenient to use equivalently
$
\kappa_2=\Lambda_{3}Q_{ab},
$
instead of $Q_{ab}$.

Let us now consider the algebra which is generated by these operators, i.e. by $\kappa_1$, $\kappa_2$ and $r$. Let $\kappa_3$ be a fourth generator defined by
\begin{align}
\kappa_3\equiv[\kappa_1,\kappa_2].
\end{align}
A direct  computation shows that:
\begin{gather}
\label{Un}
[\kappa_1,r]=0,\quad \{\kappa_2,r\}=-r+(\lambda_1+\lambda_2\lambda_3),\quad \{\kappa_3,r\}=0,\\
\label{Deux}
[\kappa_1,\kappa_3]=\kappa_2-\frac{1}{2}(\lambda_1+\lambda_2\lambda_3)r+1/2,\\
\label{Trois}
[\kappa_3,\kappa_2]=4\kappa_1+(\lambda_1+\lambda_2\lambda_3)\kappa_3r-2(\lambda_1+\lambda_2\lambda_3)\kappa_1r+\lambda_4(\lambda_1-\lambda_2\lambda_3)r.
\end{gather}
In this instance the Casimir operator for the algebra is given by
\begin{align}
Q_{C.G.}=4\kappa_1^2+\kappa_2^2-\kappa_3^2+\kappa_2-(\lambda_1+\lambda_2\lambda_3)r,
\end{align}
and acts as a multiple  $q_{C.G.}$ of the identity:
\begin{align}
q_{C.G.}=\frac{1}{4}(\lambda_1+\lambda_2\lambda_3)^2+\lambda_4^2-\frac{5}{4}.
\end{align}
It is seen that the algebra $\mathcal{H}$ of the dual $-1$ Hahn polynomials arises as the hidden symmetry algebra of the Clebsch-Gordan problem of $sl_{-1}(2)$. Indeed, redefining $K_1\rightarrow K_1+\rho/4$ in \eqref{First}, \eqref{Second} and \eqref{Third} yields an algebra of the form \eqref{Un}, \eqref{Deux} and \eqref{Trois}.

In order to establish the exact correspondence between the Clebsch-Gordan coefficients of $sl_{-1}(2)$ and the dual $-1$ Hahn polynomials encompassed by the algebra, it is necessary to determine the spectra of the operators $\kappa_1$ and $\kappa_2$. In view of the action of $A_0$ in \eqref{Product-1}, it is clear that $\kappa_1=(A_0-B_0)/2$ has a linear spectrum of the form
\begin{align}
\lambda_{\kappa_1}=n+(\mu_a-\mu_b-N)/2, \;\;\;n\in\{0,\cdots,N\}.
\end{align}
This spectrum is seen to coincide, up to a translation, with that of operator $K_1$ in the algebra $\mathcal{H}$ of the dual $-1$ Hahn polynomials.

The evaluation of the spectrum of $Q_{ab}$ is more delicate. In a given $sl_{-1}(2)$-module $(\epsilon,\mu)$, it follows from \eqref{Product-1} and \eqref{CasimirAction} that the eigenvalues of $A_0$ are given by
\begin{equation}
\label{Relation-2}
\lambda_{A_{0}}=n-\epsilon Q+1/2.
\end{equation}
In reducible representations, it is hence possible from this relation to determine the eigenvalues of the Casimir operator which are compatible with the eigenvalue $\lambda_{A_0}$ of $A_0$ that is being considered. So, for a given $\lambda_{A_0}$, the absolute value $|q|$ of the possible eigenvalues $q$ of the Casimir operator are
\begin{align}
|q|=|\lambda_{A_0}-1/2|, \,|\lambda_{A_0}-3/2|,\ldots,
\end{align}
In the calculation of the  Clebsch-Gordan coefficients of $sl_{-1}(2)$, the eigenvalue of $C_0=A_0+B_0$ is taken to be $\mu_a+\mu_b+N+1$. Consequently, the set of the absolute values of the possible eigenvalues of $Q_{c}=Q_{ab}$ given in \eqref{Casi}  is of cardinality $N+1$ and is found to be
\begin{align}
\label{admissible}
|q_{ab}|=\mu_a+\mu_b+N+1/2, \mu_a+\mu_b+N-1/2,\cdots, \mu_a+\mu_b+1/2.
\end{align}
Since the eigenvalue of $Q_{ab}$ is $-\epsilon_{ab}\mu_{ab}$, it follows that the admissible values of $\mu_{ab}$ are given by the above ensemble \eqref{admissible} with $\mu_{ab}=|q_{ab}|$.

There remains to evaluate the associated values of $\epsilon_{ab}$. To that end, consider the eigenvector $\widetilde{e}_0=e_{N}^{(\epsilon_{ab}|_{max},\mu_{ab}|_{\text{max}})}$ of the coupled basis corresponding to the maximal admissible value of $\mu_{ab}$. The state $\widetilde{e}_0$ satisfies the relations
\begin{align}
\label{Dompe}
C_0\tilde{e}_0=(\mu_a+\mu_b+N+1)\widetilde{e}_0,\qquad C_{-}\widetilde{e}_0=0.
\end{align}
On the one hand, it then follows from \eqref{Dompe} and \eqref{Product-1} that
\begin{align}
R_{c}\widetilde{e}_0=\epsilon_{ab}|_{\text{max}}\,\widetilde{e}_0.
\end{align}
On the other hand, the value of $R_{c}=R_{a}R_{b}$ is fixed to be $(-1)^{N}\epsilon_{a}\epsilon_{b}$ on the whole space so that in particular $R_{c}\widetilde{e}_0=(-1)^{N}\epsilon_{a}\epsilon_{b}\widetilde{e}_{0}$. We therefore conclude that for the state with the maximal value $\mu_{ab}=\mu_{ab}|_{\text{max}}$ of $\mu_{ab}$, the corresponding value $\epsilon_{ab}|_{\text{max}}$ of $\epsilon_{ab}$ is
\begin{align}
\epsilon_{ab}|_{max}=(-1)^{N}\epsilon_{a}\epsilon_{b}.
\end{align}
Since $Q_{ab}\widetilde{e}_{0}=Q_{ab}e_{N}^{(\epsilon_{ab}|_{\text{max}},\mu_{ab}|_{\text{max}})}=-\epsilon_{ab}|_{\text{max}}\mu_{ab}|_{\text{max}}$, it follows that this eigenvalue $q_{ab}$ of the full Casimir operator $Q_{ab}$ is 
\begin{align}
q_{ab}=(-1)^{N+1}\epsilon_{a}\epsilon_{b}(\mu_a+\mu_b+N+1/2).
\end{align}
It is easily seen that incrementing the projection from $N$ to $N+1$ adds a new eigenvalue to the set of eigenvalues of $Q_{ab}$ while preserving the admissible values of $(\epsilon_{ab},\mu_{ab})$ for the original value $N$ of the projection. Thus, by induction, the eigenvalues of the full Casimir operator $Q_{c}=Q_{ab}$ are given by
\begin{align}
q_{ab}=-\epsilon_{ab}\mu_{ab}=(-1)^{s+1}\epsilon_{a}\epsilon_{b}(\mu_a+\mu_b+s+1/2), \;\; s=0,\ldots, N.
\end{align}
It is thus seen that the spectrum of $\kappa_2$ coincide with that of $K_2$ in the algebra $\mathcal{H}$ and that the Clebsch-Gordan coefficients of $sl_{-1}(2)$ are hence proportional to the dual $-1$ Hahn polynomials.

For definiteness, let us consider the case $\epsilon_a=1=\epsilon_b$; the other cases can be treated similarly. The proportionality constant can be determined by the orthonormality condition of the Clebsch-Gordan coefficients. One has
\begin{align}
\sum_{\mu_{ab}}C_{n, N-n, N}^{\mu_a\mu_b\mu_{ab}}C_{m, N-m, N}^{\mu_a\mu_b\mu_{ab}}=\delta_{nm}.
\end{align}
By comparison of the algebras \eqref{First}, \eqref{Second} and \eqref{Third} with \eqref{Un}, \eqref{Deux} and \eqref{Trois}, there comes
\begin{align}
C_{n, N-n,N}^{\mu_a,\mu_b,\mu_{ab}}=\sqrt{\frac{\widetilde{\omega}_k}{v_n}}\,Q_{n}(z_{k};\alpha,\beta,N)
\end{align}
where
\begin{align}
\alpha=
\begin{cases}
2\mu_b+N+1, & \text{$N$ even},\\
2\mu_a, & \text{$N$ odd,}
\end{cases}
\qquad
\beta=
\begin{cases}
2\mu_a+N+1, & \text{$N$ even},\\
2\mu_b, & \text{$N$ odd,}
\end{cases}
\end{align}
\begin{align}
z_{k}=
\begin{cases}
(-1)^{k+1}(2\mu_a+2\mu_b+2k+1) , & \text{$N$ even,}\\
(-1)^{k}(2\mu_a+2\mu_b+2k+1), & \text{$N$ odd,}
\end{cases}
\qquad 
\widetilde{\omega}_k=\begin{cases}
\omega_{N-k}, & \text{$N$ even},\\
\omega_{k}, & \text{$N$ odd}
\end{cases}
\end{align}
The Clebsch-Gordan coefficients of $sl_{-1}(2)$ have thus been determined up to a phase factor by showing that the algebra underlying this problem coincides with the algebra $\mathcal{H}$ of the dual $-1$ Hahn polynomials.

\section{A ''dual'' representation of $\mathcal{H}$ by pentadiagonal matrices}
In section 2, the algebra $\mathcal{H}$ of the dual $-1$ Hahn polynomials was derived and it was shown that this algebra admits irreducible representations of dimension $N+1$ where $K_1$, $P$ are diagonal and $K_2$ is the Jacobi matrix of the dual $-1$ Hahn polynomials. We now study irreducible representations in the basis where $K_2$ is diagonal and construct the matrix elements of $K_1$ and $P$ in that basis. For the reader's convenience, we recall the defining relations of the algebra $\mathcal{H}$
\begin{gather}
\label{First-2}
[K_1,P]=0,\qquad \{K_2,P\}=-P-2\nu,\qquad \{K_3,P\}=0,\\
\label{Second-2}
[K_1,K_2]=K_3,\qquad [K_1,K_3]=K_2+\nu P+1/2,\\
\label{Third-2}
[K_3,K_2]=4 K_1+4\nu K_1P-2\nu K_3P+\sigma P+\rho,
\end{gather}
 with $P^2=\mathbb{Id}$. It is appropriate to separate the $N$ even case from the $N$ odd case. We construct the matrix elements in the $N$ odd case first.
\subsection{$N$ odd}
Consider the basis in which $K_2$ is diagonal and denote the basis vectors by $\varphi_{k}$, $k\in\{0,\ldots,N\}$. From \eqref{grid}, the eigenvalues of $K_2$ are known and given by 
\begin{align}
\lambda_{s}=(-1)^{s}\left(s+1/2+\alpha/2+\beta/2\right).
\end{align}
One sets
\begin{align}
K_2\varphi_{k}=\lambda_{k}\varphi_{k}.
\end{align}
Let $P$ have the matrix elements $M_{\ell k}$ in the basis $\varphi_{k}$ so that
$$
P\varphi_{k}=\sum_{\ell}M_{\ell k}\,\varphi_{\ell}.
$$
Consider the vector $\phi_{k}$, with $k$ fixed. Acting with the second relation of \eqref{First-2} on $\varphi_{k}$ yields
\begin{align}
\label{Rep-1}
\sum_{\ell}M_{\ell k}\left\{\lambda_{k}+\lambda_{\ell}+1\right\}\varphi_{\ell}=(\beta-\alpha)\varphi_{k},
\end{align}
where we have used the parametrization \eqref{Para-Odd}. For the term in the sum with $\ell=k$, one gets
\begin{align}
M_{2p,2p}=\frac{\beta-\alpha}{4p+2+\alpha+\beta},\qquad
M_{2p+1,2p+1}=\frac{\alpha-\beta}{4p+2+\alpha+\beta}.
\end{align}
The remainder yields
\begin{align}
\label{Rep-2}
\sum_{\ell\neq k}M_{\ell k}\left\{\lambda_{k}+\lambda_{\ell}+1\right\}\varphi_{\ell}=0.
\end{align}
Whence we must have either
\begin{align}
\label{Rep-3}
M_{\ell,k}&=0, \;\;\;\text{or}\;\;\;
\{\lambda_{k}+\lambda_{\ell}+1\}=0.
\end{align}
Since \eqref{Rep-3} is a linear equation in $\lambda_{\ell}$ (recall that $k$ is fixed), there exists one solution for each possible parity of $k$. It is seen that
\begin{align}
\label{Rep-5}
\{\lambda_{2p}+\lambda_{2p+1}+1\}&=0,
\end{align}
so that \eqref{Rep-2} is ensured for the pairs $(k=2p, \ell=2p+1)$ and $(k=2p+1,\ell=2p)$. It follows that $M_{2p+1,2p}$ and $M_{2p,2p+1}$ are arbitrary and that $M_{\ell,k}=0$ otherwise. Thus $P$ has a block-diagonal structure with $2\times 2$ blocks. By requiring that $P^2=\mathbb{Id}$, one finds that the matrices representing $K_2$ and $P$ are of the form
\begin{align}
\textstyle
K_2=
\mathrm{diag}\left(\Lambda_{0},\ldots,\Lambda_{\lfloor N/2\rfloor}\right)
&&
\textstyle
P=\mathrm{diag}\left(\Gamma_{0},\ldots,\Gamma_{\lfloor N/2\rfloor}\right)
\end{align}
where the $2\times2$ blocks have the expression
\begin{align}
\textstyle
\Lambda_{p}=
\begin{pmatrix}
\lambda_{2p} & 0\\
0 & \lambda_{2p+1}
\end{pmatrix},\qquad
\Gamma_{p}=
\begin{pmatrix}
\frac{\beta-\alpha}{4p+2+\alpha+\beta} & \frac{2(2p+1+\beta)\gamma_p}{4p+2+\alpha+\beta}\\
\frac{2(2p+1+\alpha)}{(4p+2+\alpha+\beta)\gamma_p}& \frac{\alpha-\beta}{4p+2+\alpha+\beta} 
\end{pmatrix},
\end{align}
and where the real constants $\gamma_p$, $p\in\{0,\ldots,\lfloor N/2\rfloor\}$, define a sequence of non-zero free parameters. These parameters will be treated below. Let $K_1$ have the matrix elements $N_{\ell k}$ in the basis $\varphi_{k}$ so that $$
K_1\varphi_{k}=\sum_{\ell}N_{\ell k}\varphi_{\ell}.
$$
It is seen that the commutation relation \eqref{Third-2} is equivalent to the following linear system of equations:
\begin{align}
\label{linear-1}
\sum_{\ell}&N_{\ell,2p}\left\{[\lambda_{2p}-\lambda_{\ell}]^{2}+2\nu \Gamma_p^{(1,1)}\left[(\lambda_{2p}-\lambda_{\ell})-2\right]-4\right\}\varphi_{\ell}\nonumber \\
&+\sum_{\ell}N_{\ell,2p+1}\left\{2\nu \Gamma_{p}^{(2,1)}\left[(\lambda_{2p+1}-\lambda_{\ell})-2\right]\right\}\varphi_{\ell}=
\left\{\sigma \Gamma_{p}^{(1,1)}+\rho\right\}\varphi_{2p}+\sigma\Gamma_{p}^{(2,1)}\varphi_{2p+1},\\
\sum_{\ell}& N_{\ell,2p+1}\left\{\left[\lambda_{2p+1}-\lambda_{\ell}\right]^{2}+2\nu \Gamma_{p}^{(2,2)}\left[(\lambda_{2p+1}-\lambda_{\ell})-2\right]-4\right\}\varphi_{\ell}\nonumber\\
\label{linear-2}
&+\sum_{\ell}N_{\ell,2p}\left\{2\nu\Gamma_{p}^{(1,2)}\left[(\lambda_{2p}-\lambda_{\ell})-2\right]\right\}\varphi_{\ell}=\left\{\sigma \Gamma_{p}^{(2,2)}+\rho\right\}\varphi_{2p+1}+\sigma\Gamma_{p}^{(1,2)}\varphi_{2p},
\end{align}
where $\Gamma_{p}^{(i,j)}$, $i,j\in\{1,2\}$, denotes the $(i,j)$\textsuperscript{th} component of the $p$\textsuperscript{th} block $\Gamma_{p}$. It follows from the solution of \eqref{linear-1} and \eqref{linear-2} that the matrix representing $K_1$ is block tri-diagonal:
\begin{align}
K_1=
\begin{pmatrix}
C_0 & U_1 & & & \\
D_0 & C_1 & U_2 & & \\
 & \ddots& \ddots &\ddots & \\
& & & & U_{\lfloor N/2\rfloor}\\
& & & D_{\lfloor (N-2)/2\rfloor}& C_{\lfloor N/2\rfloor}
\end{pmatrix},
\end{align}
Using the solution of the linear system \eqref{linear-1} and \eqref{linear-2} and requiring that the first relation of \eqref{First-2} and the second relation \eqref{Second-2} are satisfied yields
\footnotesize
\begin{align*}
C_p=
\begin{pmatrix}
2p-\frac{2p(N+1-2p)(2p+\alpha)}{4p+\alpha+\beta}+\frac{(2p+1)(N-2p)(2p+1+\alpha)}{4p+2+\alpha+\beta}
&
-\frac{(2p+1+\beta)(2N+2+\alpha+\beta)(\alpha+\beta)\gamma_p}{(4p+\alpha+\beta)(4p+2+\alpha+\beta)(4p+4+\alpha+\beta)}\\
-\frac{(2p+1+\alpha)(2N+2+\alpha+\beta)(\alpha+\beta)}{(4p+\alpha+\beta)(4p+2+\alpha+\beta)(4p+4+\alpha+\beta)\gamma_p} 
&
2p+1-\frac{(2p+1)(N-2p)(2p+1+\alpha)}{4p+2+\alpha+\beta}+\frac{(2p+2)(N-2p-1)(2p+2+\alpha)}{4p+4+\alpha+\beta}
\end{pmatrix}
\end{align*}
\begin{align*}
U_p=
\begin{pmatrix}
\frac{(2p-1+\beta)(N+1+2p+\alpha+\beta)\gamma_{p-1}\epsilon_p}{(4p-2+\alpha+\beta)(4p+\alpha+\beta)}
&
0\\
\frac{2(\alpha-\beta)(N+1+2p+\alpha+\beta)\epsilon_p}{(4p-2+\alpha+\beta)(4p+\alpha+\beta)(4p+2+\alpha+\beta)}
&
\frac{(2p+1+\beta)(N+1+2p+\alpha+\beta)\gamma_p\epsilon_p}{(4p+\alpha+\beta)(4p+2+\alpha+\beta)}
\end{pmatrix}
\end{align*}
\begin{align*}
D_{p}=
\begin{pmatrix}
\frac{(2p+2)(N+1-2p)(2p+1+\alpha)(2p+2+\alpha+\beta)}{(4p+2+\alpha+\beta)(4p+4+\alpha+\beta)\gamma_p\epsilon_{p+1}}
&
\frac{2(2p+2)(N-2p-1)(2p+2+\alpha+\beta)(\alpha-\beta)}{(4p+2+\alpha+\beta)(4p+4+\alpha+\beta)(4p+6+\alpha+\beta)\epsilon_{p+1}}\\
0 &
\frac{(2p+2)(N-2p-1)(2p+3+\alpha)(2p+2+\alpha+\beta)}{(4p+4+\alpha+\beta)(4p+6+\alpha+\beta)\gamma_{p+1}\epsilon_{p+1}}
\end{pmatrix}
\end{align*}
\normalsize
where the constants $\epsilon_k$, $k\in\{1,\ldots,\lfloor N/2\rfloor\}$, define a second set of non-zero free parameters. It can be checked that with their matrix elements so defined, $K_1$, $K_2$ and $P$ realize the commutation relations \eqref{First-2}, \eqref{Second-2} and \eqref{Third-2} with the Casimir eigenvalue \eqref{Casimir-Odd}. The two sequences of free parameters appearing in the representation can be reduced to one sequence $\{\theta_i\}$ by introducing the following diagonal similarity transformation
\begin{align}
\textstyle
T_{p}=
\begin{pmatrix}
\pi_{p}\,\theta_{2p} & 0\\
0 & \frac{\pi_p\theta_{2p+1}}{\gamma(p)}
\end{pmatrix}, 
&&
 \pi_{p}=\prod_{j=0}^{p-1}\frac{1}{\gamma(j)\epsilon(j+1)},
\end{align}
\normalsize
where $p\in \{0,\ldots,\lfloor N/2\rfloor\}$, is the block index and $\pi_0=1$. It should be noted that $\theta_p\neq 0$. These free parameters correspond to all the possible diagonal similarity transformations that leave the spectrum of $K_2$ and its ordering invariant. They also correspond to the freedom associated to the substitution
\begin{align}
S_{ij}=Q_{i}(x_j;\alpha,\beta,N)\rightarrow S_{ij}'=\lambda_{j}Q_{i}(x_j;\alpha,\beta,N)
\end{align}
in the transition matrix \eqref{Transition-Matrix}. They could be fixed by unitarity requirements for example. Under the transformation $T^{-1}\mathcal{O}T$, where $\mathcal{O}$ represents any element of the algebra, the matrix elements become
\footnotesize
\begin{align*}
\Lambda_{p}=
\begin{pmatrix}
 \lambda_{2p} & 0\\
0 & \lambda_{2p+1}
\end{pmatrix},
&&
\Gamma_{p}=
\begin{pmatrix}
 \frac{\beta-\alpha}{4p+2+\alpha+\beta} 	&  \frac{2(2p+1+\beta)\theta_{2p+1}}{(4p+2+\alpha+\beta)\theta_{2p}} 
\\ 
 \frac{2(2p+1+\alpha)\theta_{2p+1}}{(4p+2+\alpha+\beta)\theta_{2p}} &  \frac{\alpha-\beta}{4p+2+\alpha+\beta}
\end{pmatrix},
\end{align*}

\begin{align*}
\textstyle
C_{p}=
\begin{pmatrix}
2p-\frac{2p(N+1-2p)(2p+\alpha)}{4p+\alpha+\beta}+\frac{(2p+1)(N-2p)(2p+1+\alpha)}{4p+2+\alpha+\beta}
&
-\frac{(2p+1+\beta)(2N+2+\alpha+\beta)(\alpha+\beta)\theta_{2p+1}}{(4p+\alpha+\beta)(4p+2+\alpha+\beta)(4p+4+\alpha+\beta)\theta_{2p}}
 \\
-\frac{(2p+1+\alpha)(2N+2+\alpha+\beta)(\alpha+\beta)\theta_{2p}}{(4p+\alpha+\beta)(4p+2+\alpha+\beta)(4p+4+\alpha+\beta)\theta_{2p+1}}
 &
2p+1-\frac{(2p+1)(N-2p)(2p+1+\alpha)}{4p+2+\alpha+\beta}+\frac{(2p+2)(N-2p-1)(2p+2+\alpha)}{4p+4+\alpha+\beta}
\end{pmatrix}
\end{align*}
\begin{align*}
U_{p}=
\begin{pmatrix}
\frac{(2p-1+\beta)(N+1+2p+\alpha+\beta)\theta_{2p}}{(4p-2+\alpha+\beta)(4p+\alpha+\beta)\theta_{2p-2}}
& 
0 
\\
\frac{2(\alpha-\beta)(N+1+2p+\alpha+\beta)\theta_{2p}}{(4p-2+\alpha+\beta)(4p+\alpha+\beta)(4p+4+\alpha+\beta)\theta_{2p+1}}
&
\frac{(2p+1+\alpha+\beta)(N+1+2p+\alpha+\beta)\theta_{2p+1}}{(4p+\alpha+\beta)(4p+2+\alpha+\beta)\theta_{2p-1}}
\end{pmatrix}
\end{align*}
\begin{align*}
D_{p}=
\begin{pmatrix}
\frac{(2p+2)(N-2p-1)(2p+1+\alpha)(2p+2+\alpha+\beta)\theta_{2p}}{(4p+2+\alpha+\beta)(4p+4+\alpha+\beta)\theta_{2p+2}}
&
\frac{2(2p+2)(N-2p-1)(\alpha-\beta)(2p+2+\alpha+\beta)\theta_{2p+1}}{(4p+2+\alpha+\beta)(4p+4+\alpha+\beta)(4p+6+\alpha+\beta)\theta_{2p+2}}
 \\
0 
&
\frac{(2p+2)(N-2p-1)(2p+3+\alpha)(2p+2+\alpha+\beta)\theta_{2p+1}}{(4p+4+\alpha+\beta)(4p+6+\alpha+\beta)\theta_{2p+3}}
\end{pmatrix},
\end{align*}
\normalsize
It is thus seen that the algebra $\mathcal{H}$ admits an irreducible representation of dimension $N+1$ where the operator $K_2$ is diagonal and where $K_1$ is the five-diagonal matrix with  elements as given by the formulas above.
\subsection{$N$ even}
The treatment of the $N$ even case is similar to that of the $N$ odd case. We consider again the basis $\phi_{k}$ in which $K_2$ is diagonal with spectrum
\begin{align}
\lambda_{s}=(-1)^{s}(s+1/2-\alpha/2-\beta/2).
\end{align}
It follows from the second commutation relation of \eqref{First-2} that the matrices $K_2$ and $P$ are of the form
\begin{align}
K_2=\mathrm{diag}(\Lambda_0,\ldots,\Lambda_{(N-2)/2},\lambda_{N}),\qquad P=\mathrm{diag}(\Gamma_{0},\ldots,\Gamma_{(N-2)/2},1),
\end{align}
where
\begin{align}
\Lambda_{p}=
\begin{pmatrix}
\lambda_{2p} & 0 \\
0 & \lambda_{2p+1}
\end{pmatrix},
\qquad
\Gamma_{p}=
\begin{pmatrix}
\frac{2N+2-\alpha-\beta}{4p+2-\alpha-\beta} & \frac{2(N+2+2p+-\alpha-\beta)\gamma_p}{4p+2+\alpha+\beta} \\
\frac{2(2p-N)}{(4p+2-\alpha-\beta)\gamma_p} & -\frac{2N+2-\alpha-\beta}{4p+2-\alpha-\beta}
\end{pmatrix}.
\end{align}
Imposing the commutation relations \eqref{Third-2}, \eqref{First-2} and the last of \eqref{Second-2}, one obtains the matrix \phantom{-} elements of $K_1$ with two sets of free parameters. The two sets can be reduced to one set $\{\xi_i\}_{i=0}^{N}$, $\xi_i\neq 0$, corresponding to the possible diagonal transformations preserving the spectrum of $K_2$ as well as its ordering. One finds
\footnotesize
\begin{align*}
\Lambda_{p}=
\begin{pmatrix}
\lambda_{2p} & 0 \\
0 & \lambda_{2p+1}
\end{pmatrix},
&&
\Gamma_{p}=
\begin{pmatrix}
\frac{2N+2-\alpha-\beta}{4p+2-\alpha-\beta} & \frac{2(N+2 p+2-\alpha -\beta)\xi_{2p+1}}{ (4p+2-\alpha-\beta)\xi_{2p}} \\
\frac{2 (2p-N) \xi_{2p}}{(4p+2-\alpha-\beta)\xi_{2p+1}}& -\frac{2N+2-\alpha-\beta}{4p+2-\alpha-\beta}
\end{pmatrix}.
\end{align*}
\begin{align*}
C_p=
\begin{pmatrix}
2p+\frac{(N-2p)(2p+1)(2p+1-\alpha)}{(4p+2-\alpha-\beta)}-\frac{2p(N+1-2p)(2p-\alpha)}{(4p-\alpha-\beta)}
&
\frac{(N+2p+2-\alpha-\beta)(\alpha^2-\beta^2)\xi_{2p+1}}{(4p-\alpha-\beta)(4p+2-\alpha-\beta)(4p+4-\alpha-\beta)\xi_{2p}}
 \\
\frac{(N-2p)(\alpha^2-\beta^2)\xi_{2p}}{(4p-\alpha-\beta)(4p+2-\alpha-\beta)(4p+4-\alpha-\beta)\xi_{2p+1}}
 & 
2p+1-\frac{(N-2p)(2p+1)(2p+1-\alpha)}{(4p+2-\alpha-\beta)}+\frac{(2p+2)(N-2p-1)(2p+2-\alpha)}{(4p+4-\alpha-\beta)}
\end{pmatrix}
\end{align*}
\begin{align*}
U_{p}=
\begin{pmatrix}
\frac{2p(N+2-2p)(2p-\alpha-\beta)(2p+N-\alpha-\beta)\xi_{2p}}{(4p-2-\alpha-\beta)(4p-\alpha-\beta)\xi_{2p-2}}
& 
0
 \\
-\frac{4p(N+2-2p)(2p-\alpha-\beta)(2N+2-\alpha-\beta)\xi_{2p}}{(4p-2-\alpha-\beta)(4p-\alpha-\beta)(4p+2-\alpha-\beta)\xi_{2p-1}} 
& 
\frac{2p(N+2-2p)(2p-\alpha-\beta)(N+2p+2-\alpha-\beta)\xi_{2p+1}}{(4p-\alpha-\beta)(4p+2-\alpha-\beta)\xi_{2p-1}}
\end{pmatrix}
\end{align*}
\begin{align*}
D_{p}=
\begin{pmatrix}
\frac{(2p+2-\alpha)(2p+2-\beta)\xi_{2p}}{(4p+2-\alpha-\beta)(4p+4-\alpha-\beta)\xi_{2p+2}}
&
\frac{2(2p+2-\alpha)(2p+2-\beta)(2N+2-\alpha-\beta)\xi_{2p+1}}{(N-2p)(4p+2-\alpha-\beta)(4p+4-\alpha-\beta)(4p+6-\alpha-\beta)\xi_{2p+2}} 
\\
0 
& 
\frac{(N-2p-2)(2p+2-\alpha)(2p+2-\beta)\xi_{2p+1}}{(N-2 p) (4p+4-\alpha-\beta)(4p+6-\alpha-\beta)\xi_{2p+3}}
\end{pmatrix}
\end{align*}
\normalsize
It is straightforward to verify that this reproduces the algebra $\mathcal{H}$ with the fixed value of the Casimir operator \eqref{Casimir-Odd}.
\section{Conclusion}
In this paper, we have derived the algebra $\mathcal{H}$ associated to the dual $-1$ Hahn polynomials and shown that this algebra occurs as the hidden symmetry algebra of the Clebsch-Gordan problem of $sl_{-1}(2)$. We also obtained the irreducible representations of $\mathcal{H}$ which involve five-diagonal matrices and correspond to the difference equation of the dual -1 Hahn polynomials. Although the algebra $\mathcal{H}$ has been derived using a specific realization with a fixed value of the Casimir operator, it can also be considered in an abstract fashion. In concluding we hence wish to offer a different presentation of $\mathcal{H}$ that makes its structure transparent. Upon introducing the following new generators:
\begin{align}
\widetilde{K_1}=K_1+\rho/4, && \widetilde{K_2}=\frac{1}{2}\left(K_2+\nu P+1/2\right), &&& \widetilde{K_3}=\frac{1}{2}K_3
\end{align}
the defining relations now take the form:
\begin{gather}
[\widetilde{K_1},P]=0,\qquad \{\widetilde{K_2},P\}=0,\qquad \{\widetilde{K_3},P\}=0\\
[\widetilde{K_1},\widetilde{K_2}]=\widetilde{K_3},\quad [\widetilde{K_1},\widetilde{K_3}]=\widetilde{K_2},\\
[\widetilde{K_3},\widetilde{K_2}]=\widetilde{K_1}+\nu\widetilde{K_1}P+\chi P,
\end{gather}
where $\chi=(\sigma-\nu\rho)/4$. This presentation makes it manifest that $\mathcal{H}$ is a 2-parameter generalization of $\mathfrak{u}(2)$ (allowing for the presence of a central element possibly hidden in $\chi$) with the inclusion of the involution $P$. The Casimir operator of $\mathcal{H}$ then takes the form
\begin{align}
Q_{\mathcal{H}}=\widetilde{K}_1^{2}+\widetilde{K}_2^{2}-\widetilde{K}_3^{2}+(\nu/2)P,
\end{align}
which is clearly a simple one-parameter deformation of the standard $sl(2)$ Casimir operator.

Interestingly, one-parameter versions of this algebra (with either $\nu$ or $\chi$ equal to zero) have been introduced in studies of finite analogues of the parabosonic oscillator\cite{VDJ-2011,VDJ-2012} . The wave functions that were found in this context turn out to be symmetrized dual $-1$ Hahn polynomials. Indeed, it is seen that the substitution
\begin{align}
\widetilde{K_2}=\frac{1}{2}\left(K_2+\nu P+1/2\right),
\end{align}
corresponds to a symmetrization of the recurrence relation of the dual -1 Hahn polynomials (i.e. the suppression of the diagonal term). It is worth pointing out that recently the algebra $\mathcal{H}$ has also been identified as the symmetry algebra of a two-dimensional superintegrable model with reflections \cite{Genest-2012-2}.

The algebra $\mathcal{H}$ can be considered as an extension of the Askey-Wilson algebra $AW(3)$\cite{Zhedanov-1991} . It is known\cite{Koornwinder-2007,Koornwinder-2008} that these algebras are related to double-affine Hecke algebras (DAHA). It would be of interest in the future to explore the possible relation between DAHAs and the algebra $\mathcal{H}$.
\section*{Acknowledgements}
V.X.G. holds a scholarship from Fonds de Recherche du Qu\'ebec-Nature et technologies (FQRNT). The research of L.V. is supported in part by the Natural Sciences and Engineering Research Council of Canada (NSERC). A.Zh. wishes to thank the Centre de recherches math\'ematiques (CRM) for its hospitality.

\footnotesize
\begin{multicols}{2}
%\bibliographystyle{plain}
%\bibliography{ref}

\end{multicols}
\end{document}